\documentclass{PoS}

\usepackage{graphics,graphicx}
\usepackage{bm}
\usepackage{epsfig}
\usepackage{mathtools}
\usepackage{epstopdf}
\usepackage{slashed}
\usepackage{stackrel}
\usepackage[normalem]{ulem}

\newcommand{\lb}{\Big{\lbrack}}
\newcommand{\rb}{\Big{\rbrack}}
\newcommand{\lp}{\Big{(}}
\newcommand{\rp}{\Big{)}}

\newcommand{\nn}{\nonumber}

\newcommand{\bmat}[1]{{\boldsymbol{\mathrm{#1}}}}

\newcommand{\ve}{\lambda}

\title{Toward an effective theory of quarkonium production in nuclear matter}

\ShortTitle{Quarkonia in QCD matter}

\author{\speaker{Ivan Vitev}
\\
  Los Alamos National Laboratory, Theoretical Division, Mail Stop B283,
       Los Alamos, NM 87544 \\
        E-mail: \email{ivitev@lanl.gov}}

\abstract{These proceedings are dedicated to Miklos Gyulassy's 70$^{\rm th}$  birthday.  In his long and distinguished
career he has made seminal contributions to many areas of heavy ion theory, including early papers on quarkonium 
phenomenology in fixed-target heavy ion experiments. Theoretical and experimental studies of the $J/\psi$ and $\Upsilon$ 
states have evolved considerably in the past decades, and I describe a recent generalization of  non-relativistic Quantum Chromodynamics 
to include interactions in a generic nuclear medium. NRQCD with Glauber gluons aims to provide a universal microscopic  description of the interaction of heavy quarkonia in a range of phases that include cold nuclear matter, dense hadron gas, and quark-gluon plasma.  Such effective field theory 
  is an important  step toward understanding the  common trends in proton-nucleus  and  nucleus-nucleus  data on quarkonium suppression.    }

\FullConference{13th International Workshop in High pT Physics in the RHIC and LHC Era (High-pT2019)\\
		19-22 March 2019\\
		Knoxville, Tennessee, USA}

\begin{document}

\section{Introduction}

A significant part of  Miklos Gyulassy's research has been dedicated to the physics of jet quenching, as described in other talks in this
symposium~\cite{Wang:2019vaz}.  In the late 1980s, together with Gavin and Jackson, he co-authored a couple of paper on  
$J/\psi$ modification in fixed-target  O+U  collisions~\cite{Gavin:1988tw,Gavin:1988hs}, pointing out that quarkonium suppression 
can occur not only in the QGP but also in a hadron gas. This observation remains pertinent today, as experimental measurements 
of the suppression of excited  versus  ground bottomonium states  $\Upsilon(2S)/\Upsilon(1S)$  as a function of the number of charged particle tracks,   shows  the same trend for  high-multiplicity p+p, p+Pb, and Pb+Pb collisions at the LHC~\cite{Chatrchyan:2013nza}.  $\psi', \, \chi_c$ and $\Upsilon$ suppression  in  d+Au reactions was also measured at RHIC~\cite{Adare:2013ezl,Adamczyk:2013poh}. 

\begin{figure}[b!]
  \centerline{\includegraphics[width = 0.9 \textwidth]{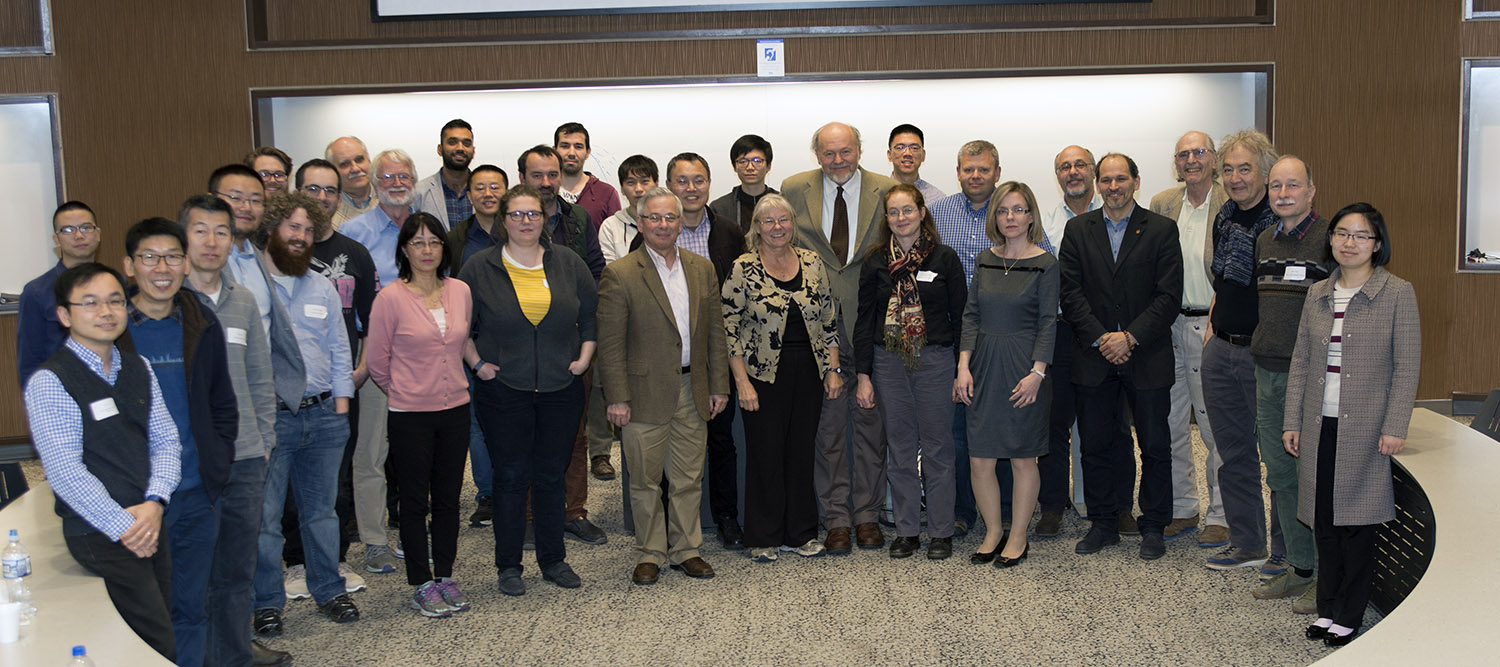}}
  \caption{Miklos Gyulassy and Gyorgyi Gyulassy with colleagues and  friends at the 13$^{\rm th}$ International Workshop in High $p_T$ Physics in the RHIC and LHC Era. }
  \label{fig:mg}
\end{figure}

To address the modification of quarkonium cross sections in  p+A  and A+A reactions,  we aim to develop a universal  microscopic theory of $J/\psi$s and $\Upsilon$s applicable to different phases of nuclear matter. The effective field theory  approach is particularly suitable, as  it can provide a model-independent description of the universal  physics of energetic particle production in the background of a QCD medium. It has been applied to light and open heavy flavor final 
states to formulate a new theory -  soft collinear effective theory with Glauber gluons (SCET${_{\rm G}}$)~\cite{Ovanesyan:2011xy,Kang:2016ofv}.

With this  in mind, we first observe that calculations of heavy quarkonium production involve hierarchies of momentum and mass scales. These scales  are $p_T$, $m_Q$, $m_Q \ve$, $m_Q \ve^2$, and $\Lambda_{\text{QCD}}$, where $p_T$ is the quarkonium transverse momentum, $m_Q$ the heavy quark mass, and $\ve$ the heavy quark-antiquark pair relative velocity. The established and most successful effective theory that describes quarkonium production and decays is non-relativistic QCD (NRQCD)~\cite{Bodwin:1994jh}. 
As  in the vacuum, production of quarkonia in nuclear matter remains a multi-scale problem.  We  have recently demonstrated that one can generalize NRQCD to incorporate interactions of the non-relativistic heavy quarks with the medium~\cite{Makris:2019ttx}. This was achieved by incorporating the  Glauber and Coulomb gluon exchanges of the heavy quarks with the quasiparticles of QCD matter.  We believe this version of NRQCD  will facilitate a much more  robust and accurate theoretical analysis of  the wealth of existing and upcoming quarkonium measurements.

In Section II I will first show that the successful jet quenching approach is challenged by the totality of the
ground and excited charmonium state measurements. In  Section III the formulation of  NRQCD$_{\rm G}$ is described.  The 
Lagrangian of the new theory to leading and subleading power is written down in Section IV. I conclude these proceedings
in Section V.

\section{Energy loss approach to quarkonium production}

Jet quenching~\cite{Wang:1991xy,Gyulassy:2003mc} is the physical mechanism behind the suppression of  high transverse  momentum particles and jets in ultrarelativistic nuclear collisions. Understandably, it has been suggested~\cite{Spousta:2016agr,Arleo:2017ntr}  that energy loss effects can reduce the cross section of $J/\psi$ production at the LHC~\cite{Khachatryan:2016ypw,Aaboud:2018quy}. In Ref.~\cite{Makris:2019ttx} we revisited 
this conjecture making use of the leading power (LP) factorization of NRQCD~\cite{Nayak:2005rw,Fleming:2012wy}, which is expected to hold at high transverse momenta ($p_T \gg m_{Q}$).  In this limit the production cross section is factorized into short distance matching coefficients that describe the production and propagation of a parton $k$ and the  NRQCD fragmentation functions,
\begin{equation}
  d\sigma_{ij \to \mathcal{Q} +X}(p_T) = \sum_n \int_{x_{\rm min}}^{1} \frac{dx}{x} d\sigma_{ij \to k + X'}\lp\frac{p_T}{x} , \mu \rp  D_{k/\mathcal{Q}}^{\;n}(x, \mu) \;.
 \label{lp} 
\end{equation}
The dependence on the factorization scale, $\mu$,  allows for the resummation of large logarithms through the use of renormalization group techniques such as the DGLAP evolution of the fragmentation functions.  The NRQCD fragmentation functions can be written in terms of the same long-distance matrix elements (LDMEs) that appear in the fixed order formulas, 
\begin{equation}
   D_{k/\mathcal{Q}}^{\;n}(x, \mu) =  \frac{ \langle\mathcal{O}^{\mathcal{Q}}(n) \rangle} {m_c^{[n]}} \;  d_{k/n}(x,\mu) \;,  
  \end{equation}
where $[n] =0 $ for S-wave and  $[n]=2$ for P-wave quarkonia. The short distance coefficients, $d_{k/n}(x,\mu)$, are functions of the fraction, $x$, of the parton energy transferred to the quarkonium state. Recent phenomenological applications to charmonia show that Eq.~(\ref{lp})  may hold to transverse momenta  $p_T  \sim 10$~GeV~\cite{Bodwin:2015iua}.  In this work we also considered  the $J/\psi$  feed-down from decays of excited quarkonium states --  $  \psi(2S):  \; \text{Br}\lb \psi(2S) \to J/\psi + X \rb = 61.4 \pm 0.6 \%$, 
 $ \chi_{c1}:  \; \text{Br}\lb \chi_{c1} \to J/\psi + \gamma \rb = 34.3 \pm 1.0 \% $,
  $\chi_{c2}:  \; \text{Br}\lb \chi_{c2} \to J/\psi + \gamma \rb = 19.0 \pm 0.5 \% $.

\begin{figure}[t]
    \centering
    \includegraphics[width=0.49\textwidth]{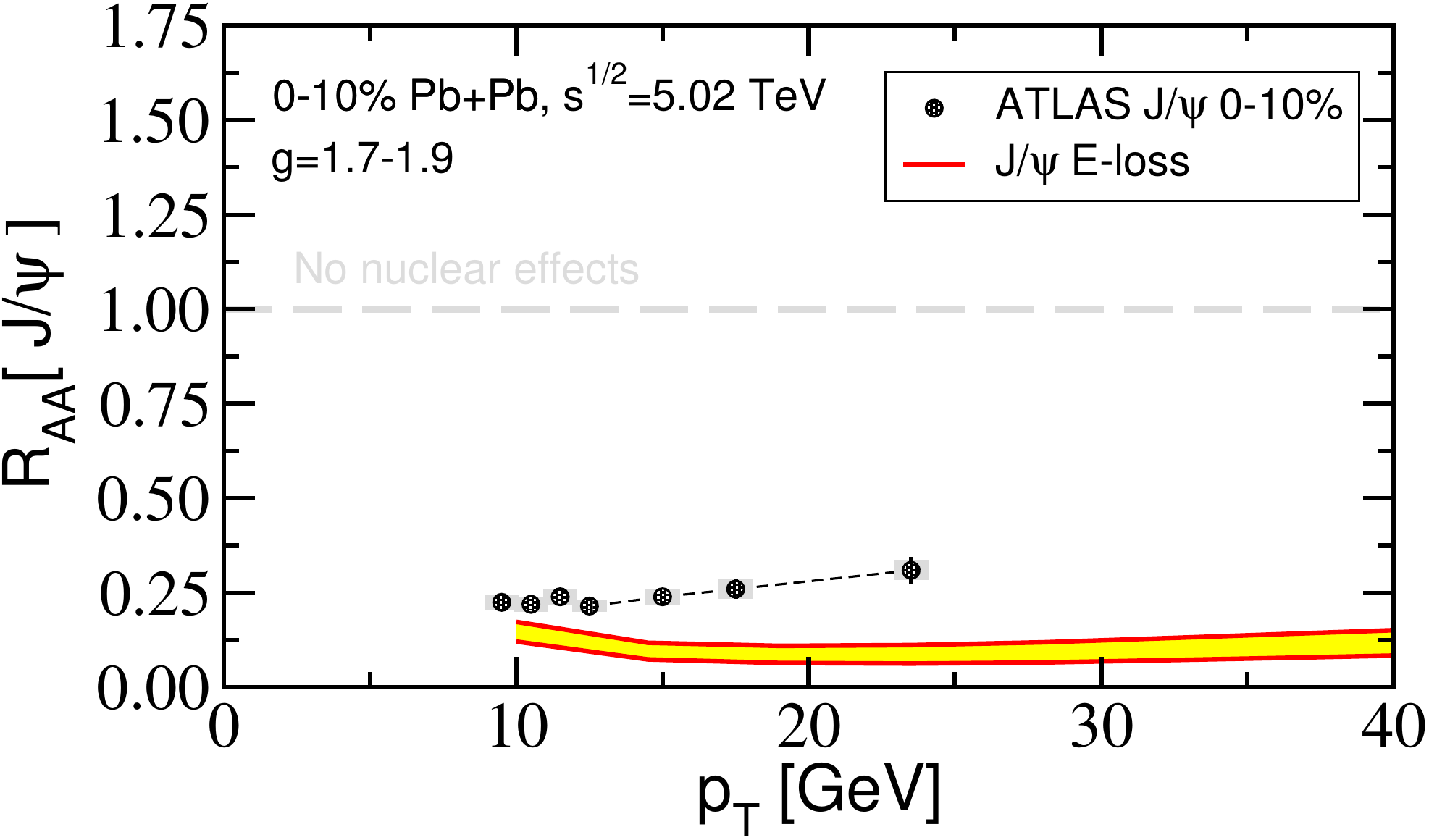}  \hspace*{0.1cm}
       \includegraphics[width=0.49\textwidth]{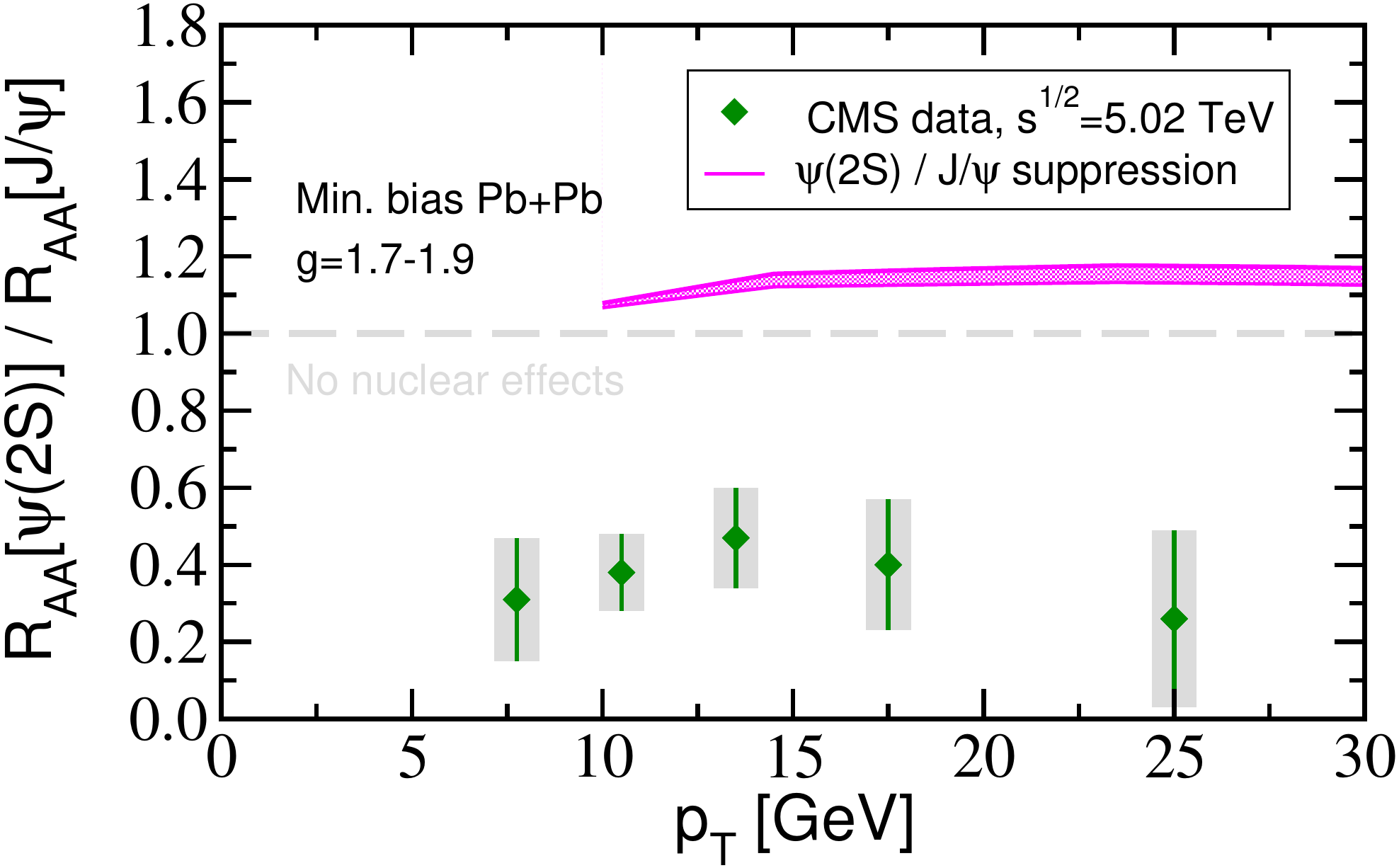}
    \vspace{-0.5cm}
    \caption{Left:  comparison of the  suppression of $J/\psi$ (yellow band) evaluated in an energy loss model with coupling between the parton and the medium $g=1.7 - 1.9$  to ATLAS data from $\sqrt{s_{NN}} = 5.02$~TeV Pb+Pb collisions at the LHC~\cite{Aaboud:2018quy}.    Right:   the double ratio of $\psi(2S)$ to $J/\psi$ suppression~\cite{Khachatryan:2016ypw}  as a measure of the relative significance of QCD matter effects on
  ground and excited states  is compared to energy loss model calculations (purple bands).  Figures are reproduced from Ref.~\cite{Makris:2019ttx}.}
    \label{fig:QqoniaEloss}
\end{figure}

Within the energy loss approach the cross section for  and quarkonium production per elementary nucleon-nucleon collision in the leading power limit can then be expressed as
\begin{eqnarray}
\frac{1}{\langle N_{\rm coll.} \rangle } \frac{d \sigma^{h}_{med}}{dyd^2p_T} & = &  \sum_c \int_{z_{\rm min}}^1 dz   \int_{0}^1 d\epsilon \, P(\epsilon)  \,   \frac{d \sigma^{c} \left(\frac{p_T}{(1-\epsilon) z  } \right) }{dyd^2p_{T_c}}   \frac{1}{(1-\epsilon)^2z^2}    D_{h/_c}( z ) \; .
\label{hspectrum-quench} 
\end{eqnarray}
Here,  $P(\epsilon)$ is the probability distribution for the hard parton $c$ to lose energy due to multiple gluon emission, 
$\frac{d \sigma^{c} (p_T) }{dyd^2p_{T_c}} $ is the hard partonic cross section, and  $ \langle N_{\rm coll.} \rangle $  is the average number of
binary nucleon-nucleon collisions.   We use the soft gluon emission limit of the full medium induced splitting function, which have been recently 
applied to light and heavy flavor jets and jet substructure~\cite{Li:2017wwc,Kang:2018wrs,Li:2018xuv}.  Comparison our theoretical calculations to ATLAS  data on the  transverse momentum dependence of $J/\Psi$ attenuation from  0-10\% central   $\sqrt{s_{NN}} = 5.02$~TeV Pb+Pb collisions at the LHC~\cite{Aaboud:2018quy} is shown in the left panel of Figure~\ref{fig:QqoniaEloss}.  The data is not described by the theoretical predictions. Energy loss calculations overpredict the suppression of $J/\psi$  and at high $p_T$   the discrepancy is as large as a factor of 3.  Even more importantly,
in the right  panel of Figure~\ref{fig:QqoniaEloss}  we show the relative medium-induced suppression of $\psi(2S)$ to $J/\psi$ in matter.
The energy loss model yields a slightly smaller suppression for the  $\psi(2S)$ state when compared to $J/\psi$. Conversely,  the experimental results show that the suppression of the weakly bound   $\psi(2S)$ is 2 to 3 times larger than that of
$J/\psi$. It is clear that the energy loss model  is incompatible with the  hierarchy of excited to ground state suppression of quarkonia in matter.

\section{Non-relativistic QCD with Glauber gluons}

In formulating a generic framework of quarkonium propagation in a variety of strongly-interacting media we are interested in the regime where matter itself might be non-perturbative, but the interaction with its  quasiparticles  can be described by perturbation theory. 
When an energetic particle traverses QCD matter, the interaction with the scattering centers of the medium is  typically mediated  by $t-$channel exchanges of off-shell gluons, called Glauber gluons.  We will call the new effective theory NRQCD with Glauber gluons, or NRQCD$_{\rm G}$. The Lagrangian of NRQCD$_{\rm G}$ can be constructed by adding to the vNRQCD Lagrangian~\cite{Rothstein:2018dzq}   the additional terms that include the interactions with quark and gluon sources through (virtual) Glauber/Coulomb gluons exchanges 
\begin{equation}
  \mathcal{L}_{\text{NRQCD}_{\rm G}} = \mathcal{L}_{\text{vNRQCD}} + \mathcal{L}_{Q-G/C} (\psi,A_{G/C}^{\mu,a}) + \mathcal{L}_{\bar{Q}-G/C} (\chi,A_{G/C}^{\mu,a})\;.
\label{nrqcdg}
\end{equation}
In Eq.~(\ref{nrqcdg}) the effective fields $A_{G/C}^{\mu,a}$ incorporate the information about the source fields, which can be collinear, static, or soft. To obtain the form and perform the power-counting of the terms in  $\mathcal{L}_{Q-G/C} (\psi,A_{G/C}^{\mu,a})$ we use three different approaches: 
\begin{itemize}
\item The background field approach where we perform a shift in the gluon field in the NRQCD Lagrangian ($A^{\mu}_{us} \to A^{\mu}_{us} +A^{\mu}_{G/C}$)  and then perform the power-counting established in Table~\ref{tb:scaling-A} to keep the leading contributions. 
\item A hybrid method, where from the full QCD diagrams for single effective Glauber/Coulomb gluon insertion with appropriate power-counting one can extract the Feynman rules. 
\item A matching method  where we expand in the power-counting parameter, $\lambda$, the full QCD diagrams describing the interactions of an incoming heavy quark and a light quark or a gluon. 
\end{itemize}
Although there are subtleties involved in the background field method,  the fact that all three approaches then give the same Lagrangian is a non-trivial test of our derivation. 

\begin{figure}[h!]
  \centerline{\includegraphics[width =  \textwidth]{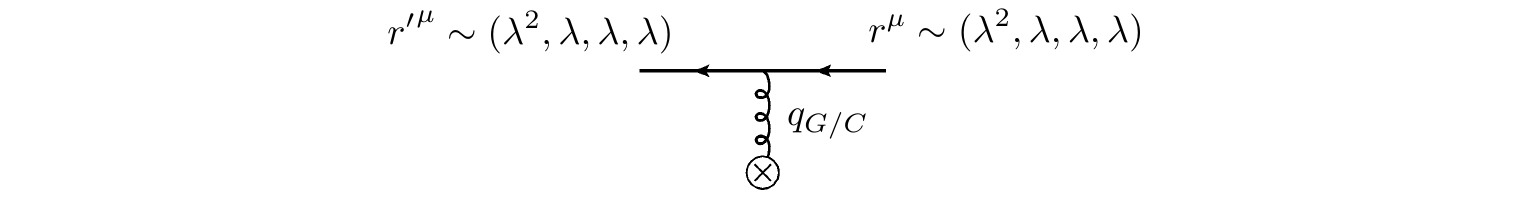}}
  \caption{Single Glauber/Coulomb gluon insertion vertex from the Lagrangian $\mathcal{L}_{Q-G/C}$, where the incoming quark caries momentum $p^{\mu} = m v^{\mu} + r^{\mu}$ and the outgoing one carries momentum ${p'}^{\mu} = m v^{\mu} + {r'}^{\mu}$.}
  \label{fig:heavy}
\end{figure}

For any gluon interacting with the vNRQCD heavy quark the scaling  $q^{0}_{G/C} \sim \lambda^2$  and $q^i_{G/C} \lesssim \lambda$ is required so that the heavy quark momenta  scale as $(\lambda^2,\bmat{\lambda})$, see Figure~\ref{fig:heavy}. If all of the three-momenta components scale as $\lambda$, i.e. $q_C^{\mu} \sim (\lambda^2,\bmat{\lambda})$ then this corresponds to Coulomb (or potential) gluons. Collinear particles cannot interact with the heavy quarks through the exchange of Coulomb gluons since this will push the collinear particles away from their canonical angular scaling. The relevant mode here is the Glauber gluons, which scale as $q_G^{\mu} \sim (\lambda^2,\lambda,\lambda,\lambda^2)$. In this EFT we include Coulomb gluons for the interaction of the heavy quarks with soft and static modes and Glauber gluons for the interactions with collinear modes:
\begin{align}
  \text{static and soft sources:}\;\;\;q^{\mu}_C &\sim (\lambda^{2},\lambda^{1},\lambda^{1},\lambda^{1})   \;, \quad 
  \text{collinear sources:}\;\;\;q^{\mu}_G &\sim (\lambda^{2},\lambda^{1},\lambda^{1},\lambda^{2})  \;.
   \label{eq:glauber-momenta}
\end{align}

The scalings of the  Glauber and Coulomb fields can be established  for different sources of scattering in the medium and are 
presented  in Table~\ref{tb:scaling-A}. Note that the scalings corresponds to the maximum allowed components for each source. 

\begin{table}[h!]
  \renewcommand{\arraystretch}{1.4}
  \begin{center}
    \begin{tabular}{|r|c|c|c|}
      \hline
      Source & Collinear      & Static       & Soft    \\ \hline \hline
      $\;\;A_C^{\mu} \sim $  &                 n.a.                       &$(\lambda^1,\lambda^2,\lambda^2,\lambda^2)$&$(\lambda^1,\lambda^1,\lambda^1,\lambda^1)$   \\ \hline
      $\;\;A_G^{\mu} \sim $  & $(\lambda^2,\lambda^3,\lambda^3,\lambda^2)$ &                 n.a.                     &               n.a.                           \\ \hline
    \end{tabular}
    \caption{The Glauber/Coulomb filed scaling for different sources of interaction in matter as calculated in Ref.~\cite{Makris:2019ttx}. }
    \label{tb:scaling-A}
  \end{center}
\end{table}

\section{The Lagrangian of NRQCD$_{\rm G}$}

The details of the derivation of the  NRQCD$_{\rm G}$ Lagrangian  using the three different methods discussed above can be 
found in our paper~\cite{Makris:2019ttx}. We summarize the  leading and subleading terms in the  Lagrangian that arise from the heavy quark sector
coupling to the medium, i.e. $\mathcal{L}_{Q-G}$, from virtual gluon insertions:
\begin{equation}
  \label{eq:L0-NR}
  \mathcal{L}_{Q-G/C}^{(0)} (\psi,A_{G/C}^{\mu,a})  = \sum_{\bmat{p},\bmat{q}_T}\psi^{\dag}_{\bmat{p}+\bmat{q}_T} \lp - g A^{0}_{G/C} \rp \psi_{\bmat{p}}\;\; (collinear/static/soft)\; , 
\end{equation}
and 
\begin{align}
  \label{eq:L1-NR}
  \mathcal{L}_{Q-G}^{(1)} (\psi,A_{G}^{\mu,a}) & =  g\sum_{\bmat{p},\bmat{q}_T} \psi^{\dag}_{\bmat{p} +\bmat{q}_T} \lp\frac{ 2  A_{G}^\bmat{n} (\bmat{n} \cdot \bmat{\mathcal{P}}) - i \lb ( \bmat{\mathcal{P}}_{\perp} \times \bmat{n}) A^{\bmat{n}}_{G}  \rb \cdot \bmat \sigma }{2m} \rp \psi_{\bmat{p}}\;\; (collinear)\; , \nn\\
  \mathcal{L}_{Q-C}^{(1)} (\psi,A_{C}^{\mu,a})  &= 0\;\; (static)\;,\nn\\
  \mathcal{L}_{Q-C}^{(1)} (\psi,A_{C}^{\mu,a}) & =  g\sum_{\bmat{p},\bmat{q}_T} \psi^{\dag}_{\bmat{p} +\bmat{q}_T} \lp\frac{ 2  \bmat{A}_{C} \cdot \bmat{\mathcal{P}} + [\bmat{\mathcal{P}} \cdot  \bmat{A}_{C} ] - i \lb  \bmat{\mathcal{P}} \times \bmat{A}_{C} \rb \cdot \bmat \sigma }{2m} \rp \psi_{\bmat{p}}\;\; (soft)\;,
\end{align}
where we use squared brackets in order to denote the region in which the label momentum operator, $\mathcal{P}^{\mu}$, acts.
If we consider the non-relativistic limit of the $t$-channel gluon exchange  diagram for a particular source, 
in addition to the above rules we obtain explicit expressions for the Glauber and Coulomb fields $A_{G}^{\mu,a}, \; A_{C}^{\mu,a}$. 
The interested reader can find those results in~\cite{Makris:2019ttx}.

\section{Conclusions}

Theory, phenomenology, and experimental measurements of quarkonia have gone a long way since the early investigation
of $J/\psi$ production in fixed target experiments at CERN and the works by Gavin, Gyulassy and Jackson.  In these proceedings I reported 
on the derivation of  the leading and sub-leading Lagrangians  of NRQCD$_{\rm G}$ for a single virtual gluon exchange~\cite{Makris:2019ttx}.  
This was achieved using three different approaches: i) the background field method, ii) a matching (with QCD) procedure, and iii) a hybrid method. 
Explicit results for the  Glauber and Coulomb fields were also obtained.  Even though  I described the formal aspects of  of NRQCD$_{\rm G}$,  it can be easily seen that the phenomenological studies of ground and excite $J/\psi$ and $\Upsilon$ states that we have done in the 
past~\cite{Sharma:2012dy,Aronson:2017ymv}  
correspond to  the leading medium correction  $\mathcal{L}_{Q-G/C}^{(0)} (\psi,A_{G/C}^{\mu,a})$  and  diagonal quarkonim state to 
quarkonium state transitions.  With the new theoretical framework at hand, such calculations can be extended rigorously to different
systems and to include medium-induced transitions from and to exited states. 

\providecommand{\href}[2]{#2}\begingroup\raggedright\endgroup


\begin{thebibliography}{10}

\bibitem{Wang:2019vaz}
X.-N. Wang, \emph{{30 years of jet quenching}},  in \emph{{13th International
  Workshop on High-pT Physics in the RHIC/LHC Era (HPT 2019) Knoxville, TN,
  USA, March 19-22, 2019}}, 2019,
  \href{https://arxiv.org/abs/1906.11998}{{\ttfamily 1906.11998}}.

\bibitem{Gavin:1988tw}
S.~Gavin and M.~Gyulassy, \emph{{Transverse Momentum Dependence of J/psi
  Production in Nuclear Collisions}},
  \href{https://doi.org/10.1016/0370-2693(88)91476-1}{\emph{Phys. Lett.}
  {\bfseries B214} (1988) 241}.

\bibitem{Gavin:1988hs}
S.~Gavin, M.~Gyulassy and A.~Jackson, \emph{{Hadronic J/psi Suppression in
  Ultrarelativistic Nuclear Collisions}},
  \href{https://doi.org/10.1016/0370-2693(88)90571-0}{\emph{Phys. Lett.}
  {\bfseries B207} (1988) 257}.

\bibitem{Chatrchyan:2013nza}
{\scshape CMS} collaboration, \emph{{Event activity dependence of Y(nS)
  production in $\sqrt{s_{NN}}$=5.02 TeV pPb and $\sqrt{s}$=2.76 TeV pp
  collisions}}, \href{https://doi.org/10.1007/JHEP04(2014)103}{\emph{JHEP}
  {\bfseries 04} (2014) 103} [\href{https://arxiv.org/abs/1312.6300}{{\ttfamily
  1312.6300}}].

\bibitem{Adare:2013ezl}
{\scshape PHENIX} collaboration, \emph{{Nuclear Modification of $??, ?_c$, and
  J/? Production in d+Au Collisions at $\sqrt{s_{NN}}$=200??GeV}},
  \href{https://doi.org/10.1103/PhysRevLett.111.202301}{\emph{Phys. Rev. Lett.}
  {\bfseries 111} (2013) 202301}
  [\href{https://arxiv.org/abs/1305.5516}{{\ttfamily 1305.5516}}].

\bibitem{Adamczyk:2013poh}
{\scshape STAR} collaboration, \emph{{Suppression of $\Upsilon$ production in
  d+Au and Au+Au collisions at $\sqrt{s_{NN}}$=200 GeV}},
  \href{https://doi.org/10.1016/j.physletb.2014.06.028,
  10.1016/j.physletb.2015.01.046}{\emph{Phys. Lett.} {\bfseries B735} (2014)
  127} [\href{https://arxiv.org/abs/1312.3675}{{\ttfamily 1312.3675}}].

\bibitem{Ovanesyan:2011xy}
G.~Ovanesyan and I.~Vitev, \emph{{An effective theory for jet propagation in
  dense QCD matter: jet broadening and medium-induced bremsstrahlung}},
  \href{https://doi.org/10.1007/JHEP06(2011)080}{\emph{JHEP} {\bfseries 06}
  (2011) 080} [\href{https://arxiv.org/abs/1103.1074}{{\ttfamily 1103.1074}}].

\bibitem{Kang:2016ofv}
Z.-B. Kang, F.~Ringer and I.~Vitev, \emph{{Effective field theory approach to
  open heavy flavor production in heavy-ion collisions}},
  \href{https://doi.org/10.1007/JHEP03(2017)146}{\emph{JHEP} {\bfseries 03}
  (2017) 146} [\href{https://arxiv.org/abs/1610.02043}{{\ttfamily
  1610.02043}}].

\bibitem{Bodwin:1994jh}
G.~T. Bodwin, E.~Braaten and G.~P. Lepage, \emph{{Rigorous QCD analysis of
  inclusive annihilation and production of heavy quarkonium}},
  \href{https://doi.org/10.1103/PhysRevD.55.5853,
  10.1103/PhysRevD.51.1125}{\emph{Phys. Rev.} {\bfseries D51} (1995) 1125}
  [\href{https://arxiv.org/abs/hep-ph/9407339}{{\ttfamily hep-ph/9407339}}].

\bibitem{Makris:2019ttx}
Y.~Makris and I.~Vitev, \emph{{An Effective Theory of Quarkonia in QCD
  Matter}},  \href{https://arxiv.org/abs/1906.04186}{{\ttfamily 1906.04186}}.

\bibitem{Wang:1991xy}
X.-N. Wang and M.~Gyulassy, \emph{{Gluon shadowing and jet quenching in A + A
  collisions at s**(1/2) = 200-GeV}},
  \href{https://doi.org/10.1103/PhysRevLett.68.1480}{\emph{Phys. Rev. Lett.}
  {\bfseries 68} (1992) 1480}.

\bibitem{Gyulassy:2003mc}
M.~Gyulassy, I.~Vitev, X.-N. Wang and B.-W. Zhang, \emph{{Jet quenching and
  radiative energy loss in dense nuclear matter}},
  \href{https://arxiv.org/abs/nucl-th/0302077}{{\ttfamily nucl-th/0302077}}.

\bibitem{Spousta:2016agr}
M.~Spousta, \emph{{On similarity of jet quenching and charmonia suppression}},
  \href{https://doi.org/10.1016/j.physletb.2017.01.041}{\emph{Phys. Lett.}
  {\bfseries B767} (2017) 10}
  [\href{https://arxiv.org/abs/1606.00903}{{\ttfamily 1606.00903}}].

\bibitem{Arleo:2017ntr}
F.~Arleo, \emph{{Quenching of Hadron Spectra in Heavy Ion Collisions at the
  LHC}}, \href{https://doi.org/10.1103/PhysRevLett.119.062302}{\emph{Phys. Rev.
  Lett.} {\bfseries 119} (2017) 062302}
  [\href{https://arxiv.org/abs/1703.10852}{{\ttfamily 1703.10852}}].

\bibitem{Khachatryan:2016ypw}
{\scshape CMS} collaboration, \emph{{Suppression and azimuthal anisotropy of
  prompt and nonprompt ${\mathrm{J}}/\psi $ production in PbPb collisions at
  $\sqrt{{s_{_{\text {NN}}}}} =2.76$ $\,\mathrm{TeV}$}},
  \href{https://doi.org/10.1140/epjc/s10052-017-4781-1}{\emph{Eur. Phys. J.}
  {\bfseries C77} (2017) 252}
  [\href{https://arxiv.org/abs/1610.00613}{{\ttfamily 1610.00613}}].

\bibitem{Aaboud:2018quy}
{\scshape ATLAS} collaboration, \emph{{Prompt and non-prompt $J/\psi $ and
  $\psi (2\mathrm {S})$ suppression at high transverse momentum in
  $5.02~\mathrm {TeV}$ Pb+Pb collisions with the ATLAS experiment}},
  \href{https://doi.org/10.1140/epjc/s10052-018-6219-9}{\emph{Eur. Phys. J.}
  {\bfseries C78} (2018) 762}
  [\href{https://arxiv.org/abs/1805.04077}{{\ttfamily 1805.04077}}].

\bibitem{Nayak:2005rw}
G.~C. Nayak, J.-W. Qiu and G.~F. Sterman, \emph{{Fragmentation, factorization
  and infrared poles in heavy quarkonium production}},
  \href{https://doi.org/10.1016/j.physletb.2005.03.031}{\emph{Phys. Lett.}
  {\bfseries B613} (2005) 45}
  [\href{https://arxiv.org/abs/hep-ph/0501235}{{\ttfamily hep-ph/0501235}}].

\bibitem{Fleming:2012wy}
S.~Fleming, A.~K. Leibovich, T.~Mehen and I.~Z. Rothstein, \emph{{The
  Systematics of Quarkonium Production at the LHC and Double Parton
  Fragmentation}},
  \href{https://doi.org/10.1103/PhysRevD.86.094012}{\emph{Phys. Rev.}
  {\bfseries D86} (2012) 094012}
  [\href{https://arxiv.org/abs/1207.2578}{{\ttfamily 1207.2578}}].

\bibitem{Bodwin:2015iua}
G.~T. Bodwin, K.-T. Chao, H.~S. Chung, U.-R. Kim, J.~Lee and Y.-Q. Ma,
  \emph{{Fragmentation contributions to hadroproduction of prompt$J/\psi$,
  $\chi_{cJ}$, and $\psi(2S)$ states}},
  \href{https://doi.org/10.1103/PhysRevD.93.034041}{\emph{Phys. Rev.}
  {\bfseries D93} (2016) 034041}
  [\href{https://arxiv.org/abs/1509.07904}{{\ttfamily 1509.07904}}].

\bibitem{Li:2017wwc}
H.~T. Li and I.~Vitev, \emph{{Inverting the mass hierarchy of jet quenching
  effects with prompt $b$-jet substructure}},
  \href{https://doi.org/10.1016/j.physletb.2019.04.052}{\emph{Phys. Lett.}
  {\bfseries B793} (2019) 259}
  [\href{https://arxiv.org/abs/1801.00008}{{\ttfamily 1801.00008}}].

\bibitem{Kang:2018wrs}
Z.-B. Kang, J.~Reiten, I.~Vitev and B.~Yoon, \emph{{Light and heavy flavor
  dijet production and dijet mass modification in heavy ion collisions}},
  \href{https://doi.org/10.1103/PhysRevD.99.034006}{\emph{Phys. Rev.}
  {\bfseries D99} (2019) 034006}
  [\href{https://arxiv.org/abs/1810.10007}{{\ttfamily 1810.10007}}].

\bibitem{Li:2018xuv}
H.~T. Li and I.~Vitev, \emph{{Inclusive heavy flavor jet production with
  semi-inclusive jet functions: from proton to heavy-ion collisions}},
  \href{https://arxiv.org/abs/1811.07905}{{\ttfamily 1811.07905}}.

\bibitem{Rothstein:2018dzq}
I.~Z. Rothstein, P.~Shrivastava and I.~W. Stewart, \emph{{Manifestly Soft Gauge
  Invariant Formulation of vNRQCD}},
  \href{https://arxiv.org/abs/1806.07398}{{\ttfamily 1806.07398}}.

\bibitem{Sharma:2012dy}
R.~Sharma and I.~Vitev, \emph{{High transverse momentum quarkonium production
  and dissociation in heavy ion collisions}},
  \href{https://doi.org/10.1103/PhysRevC.87.044905}{\emph{Phys. Rev.}
  {\bfseries C87} (2013) 044905}
  [\href{https://arxiv.org/abs/1203.0329}{{\ttfamily 1203.0329}}].

\bibitem{Aronson:2017ymv}
S.~Aronson, E.~Borras, B.~Odegard, R.~Sharma and I.~Vitev, \emph{{Collisional
  and thermal dissociation of $J/\psi$ and $\Upsilon$ states at the LHC}},
  \href{https://doi.org/10.1016/j.physletb.2018.01.038}{\emph{Phys. Lett.}
  {\bfseries B778} (2018) 384}
  [\href{https://arxiv.org/abs/1709.02372}{{\ttfamily 1709.02372}}].

\end{thebibliography}


\end{document}